\documentclass[12pt,a4paper]{article}
\usepackage{mathrsfs}
\usepackage{epsfig}
\usepackage{slashbox}
\usepackage{tabularx}
\usepackage{graphics}
\usepackage{graphicx}
\usepackage{float}

\begin{document}
\title{ LFV couplings of the extra gauge boson $Z'$ and leptonic decay and production of  pseudoscalar mesons}
\author{Chong-Xing Yue and Man-Lin Cui\\
{\small Department of Physics, Liaoning  Normal University, Dalian
116029, P. R. China}
\thanks{E-mail:cxyue@lnnu.edu.cn}}
\date{\today}

\maketitle
\begin{abstract}
Considering the constraints of the lepton flavor violating (LFV) processes $\mu \rightarrow 3e$ and $\tau\rightarrow3\mu$ on the LFV couplings $Z'\ell_{i}\ell_{j}$, in the contexts of the $E_{6}$ models, the left-right (LR) models, the "alternative" left-right (ALR) models and the 331 models, we investigate the contributions of the extra gauge boson $Z'$ to the decay rates of the processes $\ell_{i}\rightarrow\ell_{j}\nu_{\ell}\nu_{\ell}$, $\tau\rightarrow\mu P $ and $P\rightarrow \mu e$ with $P=\pi^{0},\eta$ and $\eta '$. Our numerical results show that the maximal values of the branching ratios for these  processes are not dependent on the  $Z'$ mass $M_{Z'}$ at leader order. The extra gauge boson $Z'_{X}$ predicted by the $E_{6}$ models can make the maximum value of the branching ratio $Br(\tau\rightarrow\mu\nu_{\ell}\nu_{\ell})$ reach $1.1\times10^{-7}$. All $Z'$ models considered in this paper can produce significant contributions to the process $\tau\rightarrow\mu P$. However, the value of $Br(P\rightarrow\mu e)$ is far below its corresponding experimental upper bound.

\vspace{0.5cm}

\end{abstract}
\newpage

\noindent{\bf 1. Introduction }\vspace{0.5cm}

Although most of the experimental measurements are in good agreement with the standard model (SM) predictions, there are still some unexplained  discrepancies and theoretical issues that the SM can not solve. So the SM is generally regarded as an effective realization of an underlying theory to be  yet discovered. The small but  non-vanishing neutrino masses, the hierarchy and naturalness problems provide a strong motivation for contemplating new physics (NP) beyond the SM at TeV scale, which would be in an energy range accessible at the LHC.

NP may manifest itself either  directly at high energy processes that occurred at the LHC or indirectly at lower energy processes via its effects on observable  that have been precisely  measured. Generally, the NP effects may show up at rare processes where the SM contributions are forbidden  or strongly suppressed. Therefore more theorists and experimentalists have  growing interest in the rare   decays and productions  of ordinary particles. Such studies may help one to find the NP signatures or constraint NP and provide valuable information to high energy collider experiments.

An extra gauge boson $Z'$ with heavy mass occurs in many NP models beyond the SM with extended gauge symmetry, for example see Ref.[1] and references therein. Experimentally, $Z'$ boson is going to be searched at the LHC [2, 3], although it is not conclusively discovered so far. However, stringent limits on the $Z'$ mass $M_{Z'}$ are obtained, which are still model-dependent.

Among many $Z'$ models, the most general one is the non-universal $Z'$ model, which can be realized in grand unified theories, string-inspired models, dynamical symmetry breaking models, little Higgs models, 331 models. One fundamental feature of such kind of $Z'$ models is that due to the family nonuniversal couplings or the extra fermions introduced, the extra gauge boson $Z'$ has flavor-changing fermionic couplings at the tree-level, leading many interesting phenomenological implications. For example, considering the relevant experimental data about some leptonic processes, Refs.[4, 5] have obtained the constraints on the lepton flavor violating (LFV) couplings of the boson $Z'$ to ordinary leptons and further studied their implications. In this paper, we will focus our attention on the extra gauge boson $Z'$, which is predicted by several NP models and has the tree-level LFV couplings to ordinary leptons, and consider its effects on the pure leptonic decays of the neutral scalar meson $P \rightarrow \mu e$ with $P=\pi^{0}$, $\eta$ and $\eta '$ and LFV processes $\tau\rightarrow \mu P$, $\mu\rightarrow e \nu_{\ell}\nu_{\ell}$ and $\tau\rightarrow \mu \nu_{\ell}\nu_{\ell}$ with $\ell=e, \mu$ or $\tau$. Our program is that we employ the model-dependent parameters constrained by the experimental upper limits for the LFV processes $\ell_{i}\rightarrow \ell_{j}\gamma$ and $\ell_{i}\rightarrow \ell_{j}\ell_{k}\ell_{\ell}$ to estimate the decay rates under consideration.

The rest of the paper is structured as follows: in section 2, we present the interactions of the extra gauge boson $Z'$ with fermions, including the LFV couplings, and give the constraints of the LFV processes $\mu \rightarrow 3e$ and $\tau\rightarrow3\mu$ on the $Z'$ LFV couplings to ordinary leptons in the contexts of the $E_{6}$ models, the left-right (LR) models, the "alternative" left-right (ALR) models and the 331 models. Based on the allowed LFV couplings, we calculate the contributions of the extra gauge boson $Z'$ to the decays rates of the processes $\ell_{i}\rightarrow\ell_{j}\nu_{\ell}\nu_{\ell}$, $\tau\rightarrow\mu P $ and $P\rightarrow \mu e$ with $P=\pi^{0},\eta$ and $\eta '$, in sections 3 and 4. Our conclusions and simply discussions are given in section 5.

\vspace{0.5cm} \noindent{\bf 2. Constraints on the LFV couplings $Z'\ell_{i}\ell_{j}$  }

\vspace{0.5cm}In the mass eigenstate basis, the couplings of the additional gauge boson $Z'$ to the SM fermions, including the LFV couplings, can be general written as
\begin{eqnarray}
\mathcal{L}=\bar{f_{i}}\gamma^{\mu}(g_{L}^{i}P_{L}+g_{R}^{i}P_{R})f_{i}Z'_{\mu}+\bar{\ell_{i}}(g_{L}^{ij}P_{L}+g_{R}^{ij}P_{R})\ell_{j}Z'_{\mu},
\end{eqnarray}
where $f$ and $\ell $ represent the SM fermions and charged leptons, respectively, summation over $i\neq j =1$, 2, 3 is implied, $P_{L,R}=\frac{1}{2}(1\pm\gamma_{5})$ are chiral projector operators. The left(right)-handed coupling parameter $g_{L(R)} $ should be real due to the  Hermiticity of Lagrangian $\mathcal{L}$. Considering  the goal of this paper, we do not include the flavor changing couplings of $Z'$ to the SM quarks in Eq.(1).

Many $Z'$ models can induce the LFV couplings $Z'\ell_{i}\ell_{j}$, in our analysis, we will focus our attention on the following $Z'$ models as benchmark models:

(i) The $E_{6}$ models, their symmetry breaking patterns are defined in terms of a mixing angle $\alpha$. The specific values $\alpha=0$, $\frac{\pi}{2}$ and  $ \arctan(-\sqrt{\frac{5}{3}})$ correspond to the popular scenarios $Z_{X}'$  , $Z_{\psi}'$  and $Z_{\eta}'$, respectively.

(ii) The LR model, originated from the breaking $SU(2)_{L}\times SU(2)_{R}\times U(1)_{B-L}\rightarrow SU(2)\times U(1)_{Y}\times U(1)_{LR}$ with $g_{L}=g_{R}$, and where the corresponding the $Z'$ couplings are represented  by a real parameter $\alpha_{LR}$ bounded $\sqrt{2/3}\leq\alpha_{LR}\leq\sqrt{2}$. In our calculation, we will fix $\alpha_{LR}=\sqrt{2}$, which corresponds to pure LR model.

(iii) The $Z_{ALR}'$ model based on the so-called "alternative" left-right scenario.
\begin{center}
\vspace{0.5cm}
\begin{table}

\begin{center}
\begin{tabular}{|c|c|c|c|c|}

\hline
\backslashbox{{\tiny Coupling}\kern-10cm}{{\tiny Model}\kern-0.1cm}& $E_{6}$&$LR$ & $ALR$&331 \\
\hline
$g^{u}_{L}$ & $\frac{-\cos\alpha}{2\sqrt{6}}+\frac{\sqrt{10}\sin\alpha}{12}$  & $-\frac{1}{6\sqrt{2}}$ & $\frac{1}{s_{W}\sqrt{1-2s_{W}^{2}}}(-\frac{1}{6}s_{W}^2)$ &$\frac{1}{2\sqrt{3}s_{W}\sqrt{1-\frac{4}{3}s_{W}^{2}}}(-1+\frac{4}{3}s_{W}^2)$ \\
\hline
$g^{d}_{L}$ & $\frac{-\cos\alpha}{2\sqrt6}+\frac{\sqrt10\sin\alpha}{12}$  & $-\frac{1}{6\sqrt{2}}$ & $\frac{1}{s_{W}\sqrt{1-2s_{W}^{2}}}(-\frac{1}{6}s_{W}^2)$ & $\frac{1}{2\sqrt{3}s_{W}\sqrt{1-\frac{4}{3}s_{W}^{2}}}(-1+\frac{4}{3}s_{W}^2)$ \\
\hline
$g^{u}_{R}$ & $\frac{\cos\alpha}{2\sqrt6}-\frac{\sqrt10\sin\alpha}{12}$  & $\frac{5}{6\sqrt{2}}$ & $\frac{1}{s_{W}\sqrt{1-2s_{W}^{2}}}(\frac{1}{2}-\frac{7}{6}s_{W}^2)$ &$\frac{1}{2\sqrt{3}s_{W}\sqrt{1-\frac{4}{3}s_{W}^{2}}}(\frac{4}{3}s_{W}^2)$ \\
\hline
$g^{d}_{R}$ & $\frac{-3\cos\alpha}{2\sqrt6}-\frac{\sqrt10\sin\alpha}{12}$  & $-\frac{7}{6\sqrt{2}}$ & $\frac{1}{s_{W}\sqrt{1-2s_{W}^{2}}}(\frac{1}{3}s_{W}^2)$ &$\frac{1}{2\sqrt{3}s_{W}\sqrt{1-\frac{4}{3}s_{W}^{2}}}(-\frac{2}{3\sqrt{3}}s_{W}^2)$ \\
\hline
$g^{\nu}_{L}$ & $\frac{3\cos\alpha}{2\sqrt6}+\frac{\sqrt10\sin\alpha}{12}$  & $\frac{1}{2\sqrt{2}}$ & $\frac{1}{s_{W}\sqrt{1-2s_{W}^{2}}}(-\frac{1}{2}+s_{W}^2)$ &$\frac{1}{2\sqrt{3}s_{W}\sqrt{1-\frac{4}{3}s_{W}^{2}}}(1-2 s_{W}^2)$ \\
\hline
$g^{\nu}_{R}$&$0$&$0$&$0$&$0$\\
\hline
$g^{e}_{L}$ & $\frac{3\cos\alpha}{2\sqrt6}+\frac{\sqrt10\sin\alpha}{12}$  & $\frac{1}{2\sqrt{2}}$ & $\frac{1}{s_{W}\sqrt{1-2s_{W}^{2}}}(-\frac{1}{2}+s_{W}^2)$ & $\frac{1}{2\sqrt{3}s_{W}\sqrt{1-\frac{4}{3}s_{W}^{2}}}(1-2s_{W}^2)$ \\
\hline
$g^{e}_{R}$ & $\frac{\cos\alpha}{2\sqrt6}-\frac{\sqrt10\sin\alpha}{12}$  & $-\frac{1}{4\sqrt{2}}$ & $\frac{1}{s_{W}\sqrt{1-2s_{W}^{2}}}(-\frac{1}{2}+\frac{3}{2}s_{W}^2)$ & $\frac{1}{2\sqrt{3}s_{W}\sqrt{1-\frac{4}{3}s_{W}^{2}}}(-2s_{W}^2)$ \\
\hline
\end{tabular}
 \caption{Left- and right-handed couplings of the SM fermions to the extra gauge boson \hspace*{1.2cm}$Z'$ in units of$\frac{e}{c_{W}}$, in which $s_{W}=\sin \theta_{W}$ and $c_{W}=\cos \theta_{W}$, $\theta_{W}$ is the Weinberg angle.}
\end{center}
\end{table}
\end{center}

Detailed descriptions of above $Z'$ models can be found in Ref.[1] and references therein. The flavor conserving left- and right-handed couplings $g_{L}$ and $g_{R}$ of the extra gauge boson $Z'$ to the SM fermions are shown in Table 1 [6]. As a comparison, we also include in our analysis the case of $Z_{331}'$ predicted by the 331 models [7].  The couplings of  $Z_{331}'$ to the SM fermions can be unify written as functions of the parameter $\beta$ [8]. The relevant $Z_{331}'$  couplings are also given in Table 1, where we have assumed the parameter $\beta=1/\sqrt{3}$ as numerical estimation.

In general, the LFV couplings $Z'\ell_{i}\ell_{j}$ are model-dependent. The precision  measurement data and the upper limits on some LFV processes, such as $\ell_{i}\rightarrow \ell_{j}\gamma$ and $\ell_{i}\rightarrow\ell_{j}\ell_{k}\ell_{\ell}$ can give severe constraints on these couplings. From Ref.[4], one can see that the most stringent bounds on the LFV couplings $g_{L,R}^{\mu e}$, $g_{L,R}^{\tau e}$, and $g_{L,R}^{\tau\mu}$   come from the processes  $\mu\rightarrow3e$,  $\tau\rightarrow3e$, and  $\tau\rightarrow3\mu$, respectively. So we only consider the contributions of the extra gauge boson $Z'$ predicted by the NP models considered in this paper to these LFV processes and compare with the correspond experimental upper limits.
\begin{figure}[htb]
\vspace{-0.5cm}
\begin{center}
 \epsfig{file=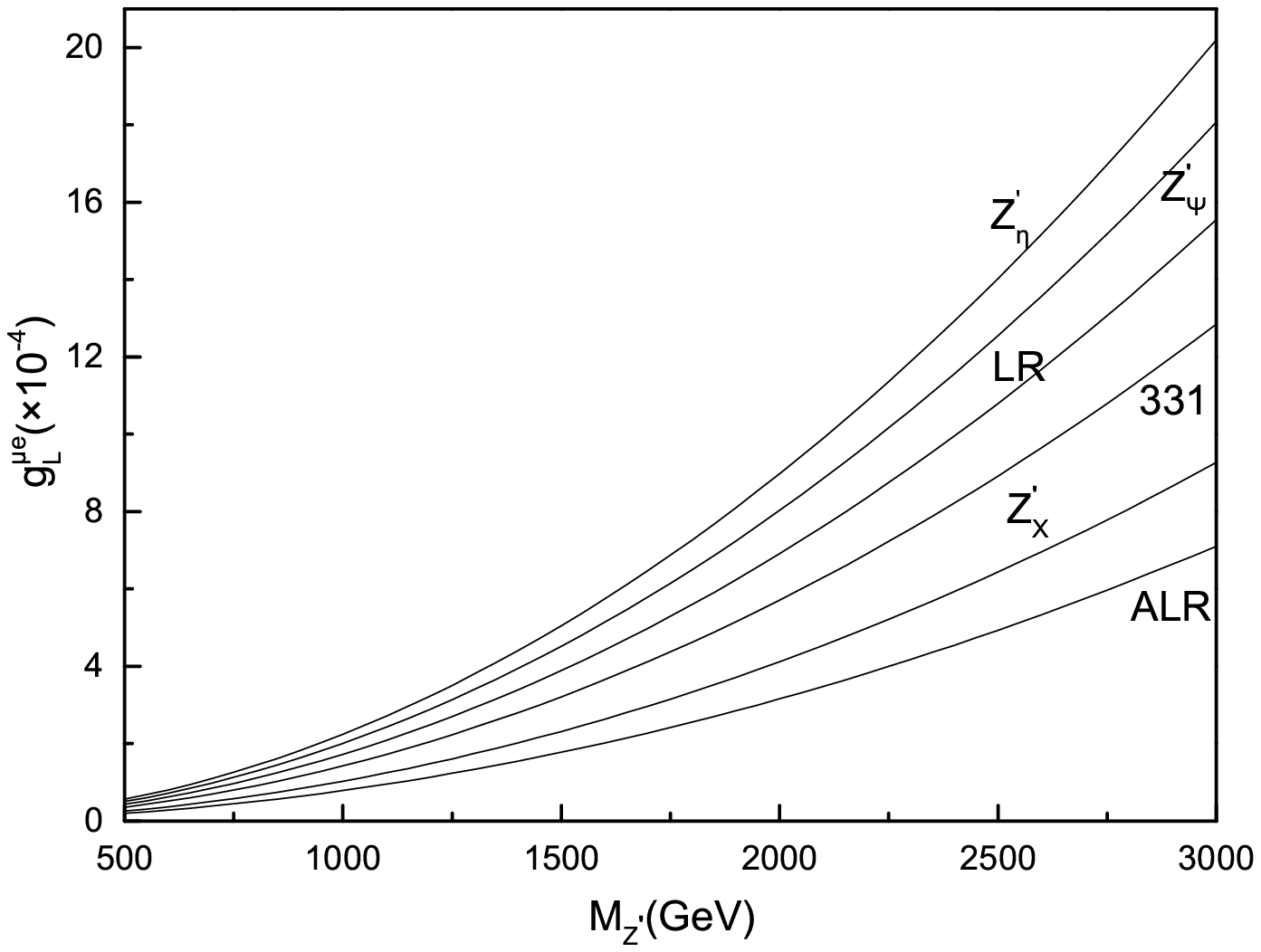, width=220pt,height=200pt}
 \epsfig{file=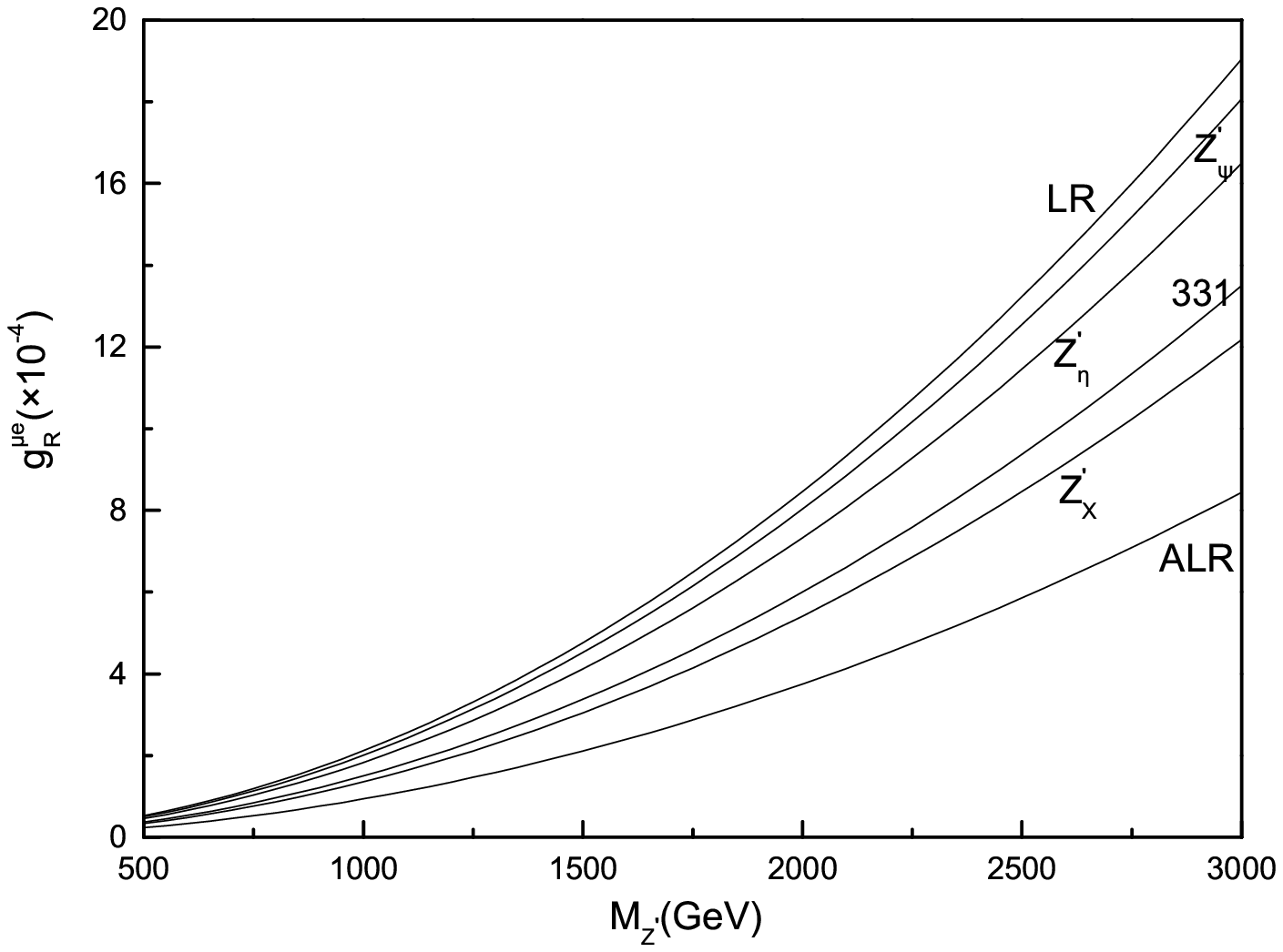, width=220pt,height=200pt}
 \vspace{-0.5cm}
 \caption{ The maximally allowed values of the left- and right-handed couplings $g_{L}^{\mu e}$ and \hspace*{1.8cm}$g_{R}^{\mu e}$ as functions of the $Z'$  mass $M_{Z'}$ for the different $Z^{'}$ models.}
 \label{ee}
\end{center}
\end{figure}

In the case of neglecting the mixing between $Z'$ and the SM $Z$, the branching ratio $Br(\ell_{i}\rightarrow\ell_{j}\ell_{j}\bar{\ell_{j}})$ can be general expressed as
\begin{eqnarray}
Br(\ell_{i}\rightarrow\ell_{j}\ell_{j}\bar{\ell_{j}})=\frac{\tau_{i}m_{i}^{5}}{1536\pi^{3}M_{Z'}^{4}}\{[2(g_{L}^{j})^{2}+(g_{R}^{j})^{2}](g_{L}^{ij})^{2}+[(g_{L}^{j})^{2}
+2(g_{R}^{j})^{2}](g_{R}^{ij})^{2}\},
\end{eqnarray}
where $\tau_{i}$ and $m_{i}$ are the lifetime and mass  of the charged lepton  $\ell_{i}$, $M_{Z'}$  is the $Z'$ mass. In above equation, we have ignored the masses of the final state leptons. In our following numerical calculation, we will take $s_{W}^{2}=0.231$, $\tau_{\tau}=4.414\times10^{11}GeV^{-1}$, $\tau_{\mu}=3.338\times10^{18}GeV^{-1}$,  $m_{\tau}=1.777GeV$  and $m_{\mu}=0.106GeV$ [9], and assume that $M_{Z'}$ is in the range of $1TeV\sim3TeV$.

\begin{figure}[htb]
\vspace{-0.5cm}
\begin{center}
 \epsfig{file=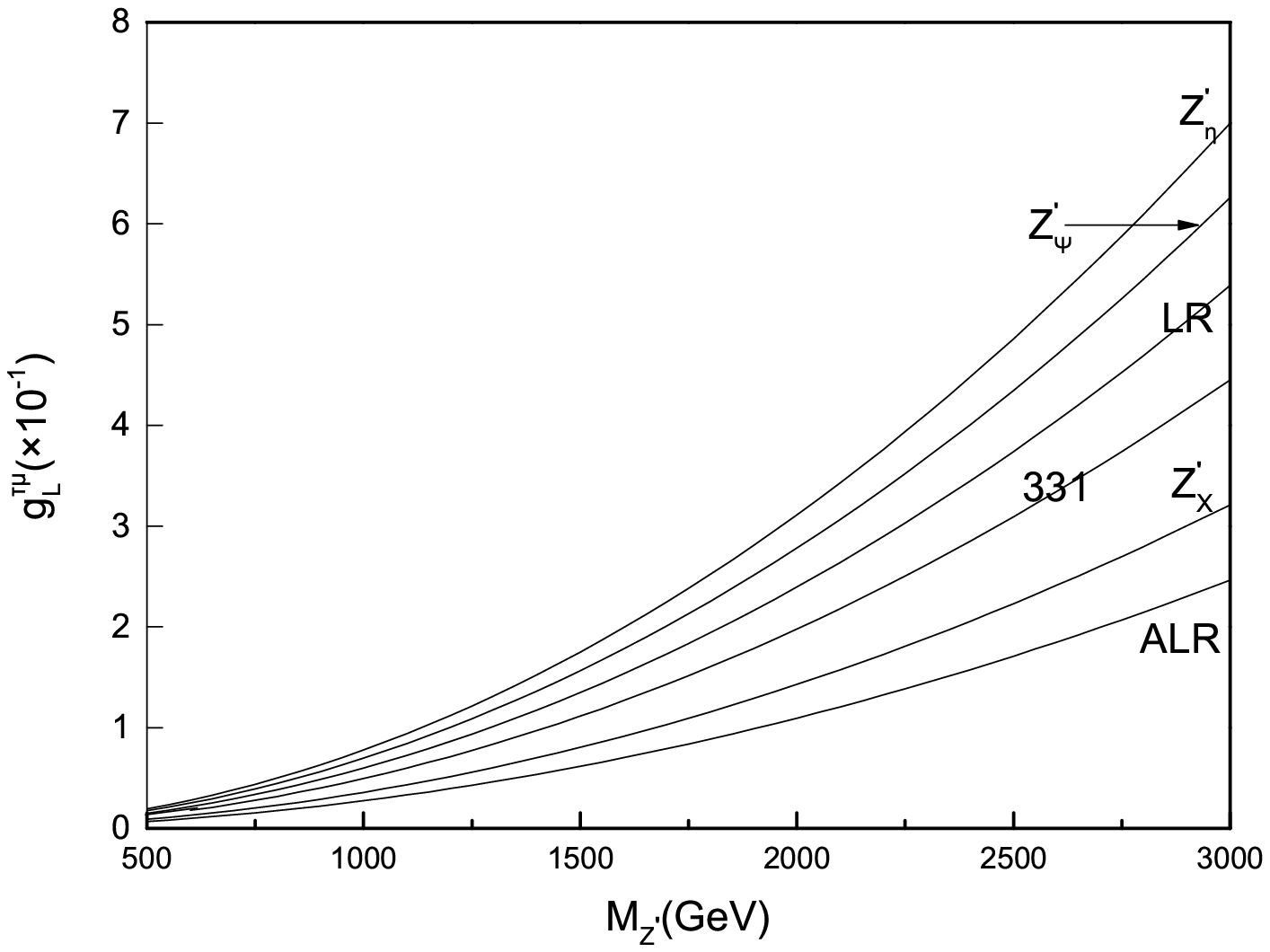, width=220pt,height=200pt}
 \epsfig{file=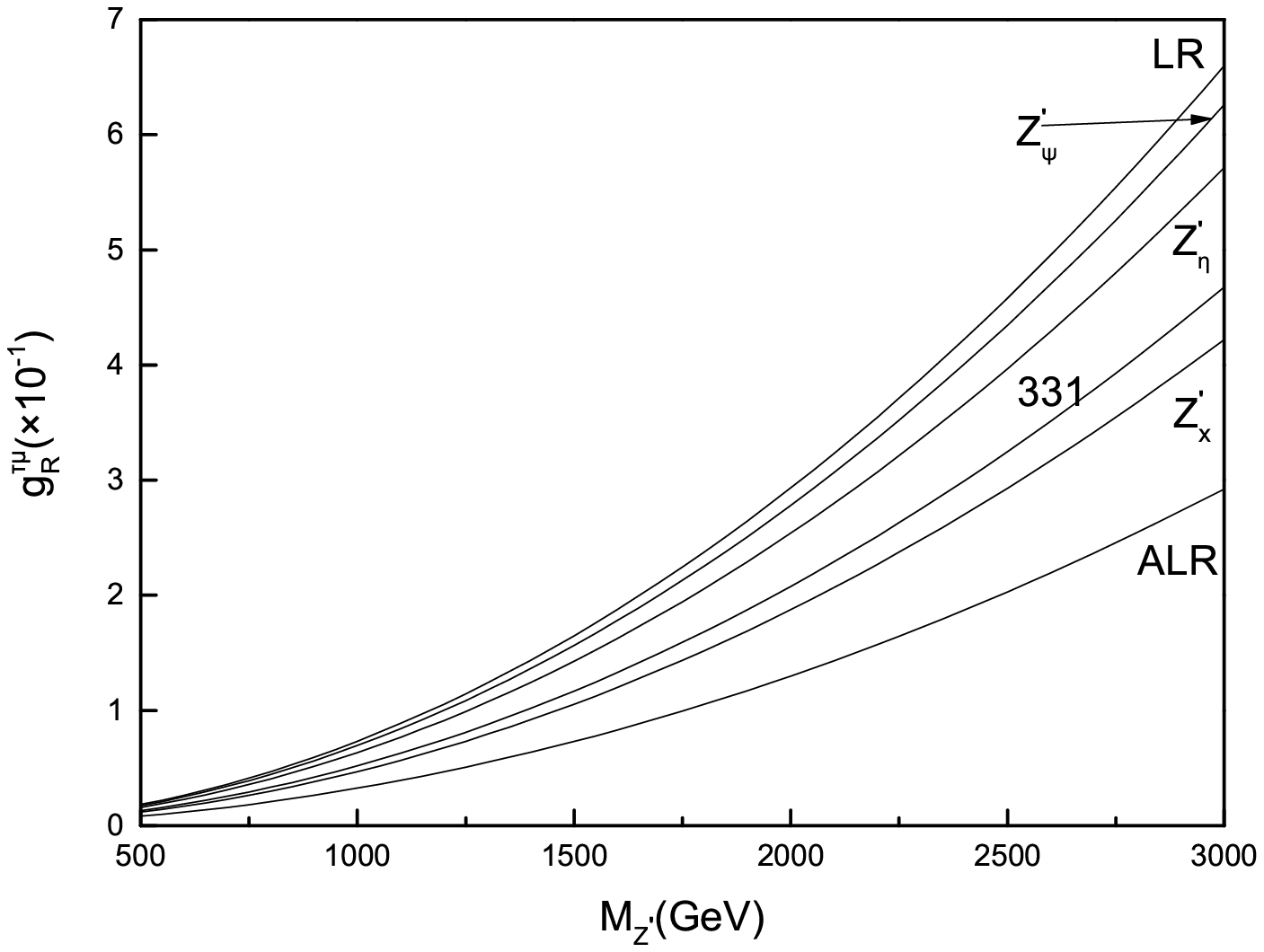, width=220pt,height=200pt}
 \vspace{-0.5cm}
  \caption{  The maximally allowed values of the  left- and right-handed couplings $g_{L}^{\tau\mu}$ and \hspace*{1.8cm}$g_{R}^{\tau\mu}$ as functions of the $Z'$  mass $M_{Z'}$ for the different $Z^{'}$ models.}
 \label{ee}
\end{center}
\end{figure}

Assuming that only of $g_{L,R}^{ij}$ is nonzero at a time, we can obtain constraints on  $g_{L,R}^{ij}$  from the current experimental upper limits [9]
\begin{eqnarray}
 Br^{exp}(\mu\rightarrow ee\bar{e})<1.0 \times 10^{-12},&& \nonumber
 Br^{exp}(\tau\rightarrow ee\bar{e})<2.7 \times 10^{-8},\\
 Br^{exp}(\tau\rightarrow \mu\mu\bar{\mu})<2.1 \times 10^{-8}.
\end{eqnarray}

Our numerical results for  different  $Z'$ models are summarized in Fig.1 and Fig.2, in which we plot the maximally allowed values of the left(right)-handed couplings $g_{L,R}^{\mu e}$ and $g_{L,R}^{\tau\mu}$ as functions of the $Z'$ mass $M_{Z'}$. One can see from these figures that the values of $g_{L,R}^{ij}$ are slight different for various $Z'$ models. The maximal values of $g_{L,R}^{\mu e}$ are far smaller than those of $g_{L,R}^{\tau\mu}$. In the following sections we will use these results to estimate the contributions of $Z'$ to the processes $\tau\rightarrow \mu\nu_{\ell}\nu_{\ell}$, $\mu \rightarrow e \nu_{\ell}\nu_{\ell}$ with $\ell=e, \mu$ or $\tau$, $P\rightarrow \mu e$, and $\tau\rightarrow\mu P$ with $P=\pi^{0}$, $\eta$ and $\eta'$.

\vspace{0.5cm} \noindent{\bf 3. The extra gauge boson $Z'$ and the LFV process $\ell_{i}\rightarrow\ell_{j}\nu_{\ell}\nu_{\ell}$\hspace*{0.6cm}}
\vspace{0.5cm}

Neutrino oscillation experiments have shown very well that neutrinos have masses and mix each other [9, 10]. Recently, the T2K experiment has confirmed the neutrino oscillation in $\nu_{\mu}\rightarrow\nu_{e}$ appearance events [11]. Thus neutrino physics is now entering a new precise measurement era.
\begin{table}[h]
\begin{center}

\begin{tabular}{|c|c|c|}

\hline
$Models$ &$Br(\tau\rightarrow\mu \nu_{\ell}\nu_{\ell}) $ &$Br(\mu\rightarrow e\nu_{\ell}\nu_{\ell})$\\
\hline
$Z_{X}^{'}$ & $1.1\times10^{-7}$  & $5.3\times10^{-12}$ \\
\hline
$Z_{\psi}^{'}$&$6.3\times10^{-8}$ &$3.0\times10^{-12}$ \\
\hline
$Z_{\eta}^{'}$&$2.8\times10^{-8}$ &$1.3\times10^{-12}$ \\
\hline
$LR$ &$9.8\times10^{-8}$&$4.7\times10^{-12}$ \\
\hline
$ALR$&$9.2\times10^{-8}$&$4.4\times10^{-12}$ \\
\hline
331&$7.1\times10^{-8}$&$3.4\times10^{-12}$ \\

\hline

\end{tabular}
 \caption{ The maximum values of the branching ratios $Br(\ell_{i}\rightarrow\ell_{j}\nu_{\ell}\nu_{\ell})$ for various $Z'$ \hspace*{1.8cm}models.}
\end{center}
\end{table}

Although there are not the relevant experimental data so far, the neutrino data allow the existence of the LFV processes $\ell_{i}\rightarrow\ell_{j}\nu_{\ell}\nu_{\ell}$ with $\ell=e, \mu$ or $\tau$, which have been studied in specific NP models [12]. From discussions given in section 2, we can see that the extra gauge boson $Z'$ can contribute these LFV processes at tree level. The branching ratios can be approximately written as
\begin{eqnarray}
Br(\tau\rightarrow\mu\nu_{\ell}\nu_{\ell})=\sum_{\ell=e,\mu,\tau}\frac{\tau_{\tau}m_{\tau}^{5}}{1536\pi^{3}M_{Z'}^{4}}(g_{L}^{\nu_{\ell}})^{2}[2(g_{L}^{\tau\mu})^{2}+
(g_{R}^{\tau\mu})^{2}],\\
Br(\mu\rightarrow e\nu_{\ell}\nu_{\ell})=\sum_{\ell=e,\mu,\tau}\frac{\tau_{\mu}m_{\mu}^{5}}{1536\pi^{3}M_{Z'}^{4}}(g_{L}^{\nu_{\ell}})^{2}[2(g_{L}^{\mu e})^{2}+(g_{R}^{\mu e})^{2}].
\end{eqnarray}

Comparing Eq.(1) with above equations one can see that, if we fix the values of the LFV couplings $Z'\ell_{i}\ell_{j}$ as the maximum values arose from the current experimental upper limits for the process $\ell_{i}\rightarrow\ell_{j}\ell_{j}\bar{\ell_{j}}$,  the branching ratios $Br(\ell_{i}\rightarrow\ell_{j}\nu_{\ell}\nu_{\ell})$ do not depend on the $Z'$ mass  $M_{Z'}$, their values differ from each other for various $Z'$ models. Our numerical results are given in Table 2. One can see from Table 2 that the maximum values of $Br(\tau\rightarrow\mu\nu_{\ell}\nu_{\ell})$ are larger than those for $Br(\mu\rightarrow e\nu_{\ell}\nu_{\ell})$ at least four orders of magnitude in the context of these $Z'$ models. For the $Z'_{X}$ model, the maximum value of $Br(\tau\rightarrow\mu\nu_{\ell}\nu_{\ell})$ can reach $1.1\times10^{-7}$.

If the decay processes $\ell_{i}\rightarrow\ell_{j}\nu_{\ell}\nu_{\ell}$ would be accurate measured in future, it might be used to test different $Z'$ model.

\vspace{0.5cm} \noindent{\bf 4. The extra gauge boson $Z'$ and  the LFV processes $\tau\rightarrow\mu P$ and $P\rightarrow\mu e$ with \hspace*{0.6cm}$P=\pi$ , $\eta$ and $\eta'$}

\vspace{0.5cm}

The LFV processes $\tau\rightarrow\mu P$ with $P=\pi$ , $\eta$ or $\eta'$ and $P\rightarrow\mu e$ are severely suppressed in the SM, which are sensitive to NP effects, for example see Refs.[13, 14, 15]. Although these processes have not been observed so far, their current experimental upper bounds have existed [9, 16]
\begin{eqnarray}
 Br(\tau\rightarrow\mu\pi)<1.1 \times 10^{-7},&& \nonumber
 Br(\tau\rightarrow\mu \eta)<6.5 \times 10^{-8},\\
 Br(\tau\rightarrow\mu\eta')<1.3 \times 10^{-7},&& \nonumber
 Br(\pi\rightarrow\mu e)<3.6 \times 10^{-10},\\
 Br(\eta\rightarrow\mu e)<6.0 \times 10^{-6},&&
 Br(\eta'\rightarrow\mu e)<4.7 \times 10^{-4}.
\end{eqnarray}

It is obvious that the LFV processes $\tau\rightarrow\mu P$ and $P\rightarrow\mu e$ can be induced at tree level by the extra  gauge boson $Z'$ considered in this paper. In the local four-fermion approximation, the effective Hamiltonian is given by
\begin{eqnarray}
\mathcal{H}=\frac{4G_{F}}{\sqrt{2}}(\frac{s_{W}M_{Z}}{M_{Z'}})^{2}[g^{\tau\mu}_{L}(\bar{\mu}\gamma^{\mu}P_{L}\tau)+g^{\tau\mu}_{R}(\bar{\mu}\gamma^{\mu}P_{R}\tau)]
\sum_{q}[g^{q}_{L}(\bar{q}\gamma_{\mu}P_{L}q)+g^{q}_{R}(\bar{q}\gamma_{\mu}P_{R}q)].
\end{eqnarray}

The relevant hadronic matrix elements that will enter in our calculations are the following [17]
\begin{eqnarray}
<P(p)\mid \overline{q}\gamma_{\mu}\gamma_{5} q\mid0>=-ib^{p}_{q}f^{q}_{p}p_{\mu},
\end{eqnarray}
where $b^{p}_{q}$ is the  form factor, $f^{q}_{p}$ is the decay constant of the corresponding meson. For the meson $\pi^{0}$, there are $q=u$ or $d$, $b^{\pi}_{u}=-b^{\pi}_{d}=1/\sqrt{2}$,
$f^{u}_{\pi}=f^{d}_{\pi}=130.4\pm0.2$MeV. For the mesons $\eta$ and $\eta'$, there are $q=u, d$ or $s$, $b^{\eta,\eta'}_{u}=b^{\eta,\eta'}_{d}=1/\sqrt{2}$,
$b^{\eta,\eta'}_{s}=1$, $f^{u}_{\eta}=f^{d}_{\eta}=108\pm3$ MeV, $f^{u}_{\eta'}=f^{d}_{\eta'}=89\pm3$MeV, $f^{s}_{\eta}=-111\pm6$MeV and $f^{s}_{\eta'}=136\pm6$MeV.

Neglecting terms of the order $O$ $(m_{\mu}/m_{\tau})$, the decay widths for the LFV decays $\tau\rightarrow\mu P$ with  $P=\pi$, $\eta$, and $\eta'$ can be approximately written as
\begin{eqnarray}
\Gamma(\tau \rightarrow \mu P)=\frac{m^{3}_{\tau}}{16\pi}(L^{2}_{P}+R^{2}_{P})(1-\frac{M^{2}_{P}}{m^{2}_{\tau}})^{2}
\end{eqnarray}
with
\begin{eqnarray}
L_{P}=\frac{4G_{F}}{\sqrt{2}}(\frac{s_{W}M_{Z}}{M_{Z'}})^{2}g^{\tau\mu}_{L}[\sum_{q}(g^{q}_{L}-g^{q}_{R})b^{p}_{q}f^{q}_{p}],
\end{eqnarray}
\begin{eqnarray}
R_{P}=\frac{4G_{F}}{\sqrt{2}}(\frac{s_{W}M_{Z}}{M_{Z'}})^{2}g^{\tau\mu}_{R}[\sum_{q}(g^{q}_{L}-g^{q}_{R})b^{p}_{q}f^{q}_{p}].
\end{eqnarray}
\begin{table}[h]
Where $M_{P}$ is the mass of the neutral pseudo-scalar meson which are taken as 134.98 MeV, 547.8 MeV and 957.78 MeV for the mesons $\pi^{0}$, $\eta$ and $\eta'$, respectively [9]. For the LFV left- and right-couplings $g_{L}^{\tau\mu}$ and $g_{R}^{\tau\mu}$, same as section 3, we also take their maximum values satisfying the current experimental upper limit for the LFV process $\tau\rightarrow3\mu$. Then one can easily obtain the maximal values of the branching ratios $Br(\tau\rightarrow\mu P)$, which are shown in Table 3 for different $Z'$ models. Among these $Z'$ models, the contribution of  $Z'$ predicted by the pure LR model to the LFV decay $\tau\rightarrow\mu P$ is the maximum. However, its maximal value of the branching ratio $Br(\tau\rightarrow\mu P)$ is still lower than the corresponding experimental upper limit at least by one order of magnitude. So, comparing with  the LFV process $\tau\rightarrow\mu P$, the LFV process $\tau\rightarrow3\mu$ can give more serve constraints on these  $Z'$ models.

\begin{center}
\begin{tabular}{|c|c|c|c|}
\hline
$Models$ &$Br(\tau\rightarrow\mu \pi) $ &$Br(\tau\rightarrow\mu \eta)$&$Br(\tau\rightarrow\mu \eta')$ \\
\hline
$Z_{X}^{'}$ & $5.1\times10^{-10}$  & $1.4\times10^{-10}$  &$1.4\times10^{-10}$ \\
\hline
$Z_{\psi}^{'}$&$0$ &$9.7\times10^{-11}$ &$2.3\times10^{-9}$ \\
\hline
$Z_{\eta}^{'}$&$5.6\times10^{-10}$ &$4.2\times10^{-10}$ & $7.2\times10^{-10}$ \\
\hline
$LR$ &$4.0\times10^{-9}$&$1.1\times10^{-9}$&$1.1\times10^{-9}$\\
\hline
$ALR$&$7.6\times10^{-11}$&$8.2\times10^{-11}$&$2.9\times10^{-10}$\\
\hline
331&$9.5\times10^{-11}$&$1.7\times10^{-10}$&$1.1\times10^{-9}$\\
\hline
\end{tabular}
 \caption{The maximal values of the branching ratios $Br(\tau\rightarrow\mu P)$ with $P=\pi^{0}$, $\eta$ and  $\eta'$, \hspace*{1.8cm}for different $Z'$ models.}
\end{center}
\end{table}

The general expression of the branching ratio $Br(P\rightarrow\ell_{i}\ell_{j})$ contributed by the extra gauge boson $Z'$ can be written as
\vspace{1.0cm}
\begin{eqnarray}
Br(P\rightarrow\ell_{i}\ell_{j})&=&\frac{G_{F}^{2}s_{W}^{4}M_{Z}^{4}}{4\pi M_{Z'}^{4}}f(x_{i}^{2},x_{j}^{2})M_{p}\tau_{p}(m_{i}+m_{j})^{2}\nonumber\\
&&\left.\mid\sum_{p}b_{q}f_{p}^{q}(g_{L}^{q}-g_{R}^{q})\mid^{2}\{(g_{L}^{ij}-g_{R}^{ij})^{2}[1-(x_{i}-x_{j})^{2}]\right.\nonumber\\
&&\left.+\frac{(m_{i}-m_{j})^{2}}{(m_{i}+m_{j})^{2}}(g_{L}^{ij}+g_{R}^{ij})^{2}[1-(x_{i}+x_{j})^{2}]\}\right.\nonumber\\
\end{eqnarray}
with
\begin{eqnarray}
f(x_{i}^{2},x_{j}^{2})=\sqrt{1-2(x_{i}^{2}+x_{j}^{2})+(x_{i}^{2}-x_{j}^{2})^{2}}, \hspace*{1.0cm} x_{i}=\frac{m_{\ell_{i}}}{M_{P}}.
\end{eqnarray}

\begin{table}[htb]

\begin{center}
\begin{tabular}{|c|c|c|c|}
\hline
$ $ &$Br(\pi\rightarrow\mu e) $ &$Br(\eta\rightarrow\mu e)$&$Br(\eta'\rightarrow\mu e)$ \\
\hline
$EXP. $&$3.6\times10^{-10}$&$6.0\times10^{-6}$&$4.7\times10^{-4}$ \\
\hline
$Models$ &$ $&$ $&$ $\\
\hline
$Z_{X}^{'}$ & $4.2\times10^{-20}$  & $1.1\times10^{-19}$  &$1.1\times10^{-19}$ \\
\hline
$Z_{\psi}^{'}$&$1.2\times10^{-19}$ &$3.0\times10^{-19}$ &$3.1\times10^{-19}$ \\
\hline
$Z_{\eta}^{'}$&$1.2\times10^{-19}$ &$3.1\times10^{-19}$ & $3.2\times10^{-19}$ \\
\hline
$LR$ &$1.1\times10^{-19}$&$1.5\times10^{-19}$&$2.9\times10^{-19}$\\
\hline
$ALR$&$2.2\times10^{-20}$&$5.6\times10^{-20}$&$5.8\times10^{-20}$\\
\hline
331&$6.3\times10^{-20}$&$1.6\times10^{-19}$&$1.7\times10^{-19}$\\
\hline
\end{tabular}
 \caption{The maximal values of the branching ratios $Br(P\rightarrow\mu e)$ with $P=\pi^{0}$, $\eta$ \hspace*{1.8cm}and  $\eta'$ for different $Z'$ models. The second row represents the corresponding \hspace*{1.8cm}experimental upper bound.}
\end{center}
\end{table}

In the contexts of the various $Z'$ models considered in this paper, using above formula, we can estimate the maximal value of the branching ratios for the LFV meson decays $\pi\rightarrow\mu e$, $\eta\rightarrow\mu e$ and $\eta'\rightarrow\mu e$. Our numerical results are given in Table 4, in which we also give the corresponding experimental upper bound. One can see from this table that, considering the constraints of the experimental upper bound for the LFV  process  $\mu\rightarrow3e$ on the LFV couplings $Z'\ell_{i}\ell_{j}$, the contributions of the extra gauge boson $Z'$ to the LFV meson decays $P\rightarrow\mu e$ are very small. The value of the branching ratio $Br(P\rightarrow\mu e)$ is far below its corresponding experimental upper bound for all of the $Z'$ models considered in this paper.

 Certainly, the extra $Z'$ also has contributions to the FC meson decays $\pi\rightarrow e^{+}e^{-}$, $\eta\rightarrow e^{+}e^{-}$ and $\mu^{+}\mu^{-}$, and $\eta'\rightarrow e^{+}e^{-}$ and $\mu^{+}\mu^{-}$. Although these decay processes are not depressed by the LFV couplings, the contributions of the extra gauge boson $Z'$ are also very small being large $Z'$ mass $M_{Z'}$. We do not show the numerical results here.

\vspace{0.5cm} \noindent{\bf 5. Conclusions and discussions }

\vspace{0.5cm}
 Many NP models beyond the SM predict the existence of the extra gauge boson $Z'$, which can induce the LFV couplings to the SM leptons at the tree-level. This kind of new particles can produce rich LFV phenomenology in current or future high energy collider experiments, which should be carefully studied. It is helpful to search for NP  models beyond the SM and further to test the SM.

 In this paper, we first consider the constraints of the experimental upper limits for the LFV processes $\ell_{i}\rightarrow \ell_{j}\gamma$ and $\ell_{i}\rightarrow \ell_{j}\ell_{k}\ell_{\ell}$ on the  LFV couplings of the extra gauge boson $Z'$ to ordinary leptons in the contexts of the $E_{6}$ models, the LR models, the ALR model and the 331 models. The most stringent bounds on the LFV couplings $g_{L,R}^{\mu e}$ and $g_{L,R}^{\tau\mu}$   come from the processes  $\mu\rightarrow3e$ and  $\tau\rightarrow3\mu$, respectively. We find that the values of $g_{L,R}^{ij}$ are slight different for various $Z'$ models. The maximal values of $g_{L,R}^{\mu e}$ are much smaller than those of $g_{L,R}^{\tau\mu}$. Then, considering these constraints, we calculate the contributions of $Z'$ to the LFV processes $\tau\rightarrow \mu\nu_{\ell}\nu_{\ell}$, $\mu \rightarrow e \nu_{\ell}\nu_{\ell}$ with $\ell=e, \mu$ or $\tau$, $P\rightarrow \mu e$, and $\tau\rightarrow\mu P$ with $P=\pi^{0}$, $\eta$ and $\eta'$ in these $Z'$ models. Our numerical results show that the maximal values of the branching ratios for these LFV decay processes are not dependent on the extra gauge boson $Z'$ mass $M_{Z'}$ at leader order. For the process $\tau\rightarrow \mu\nu_{\ell}\nu_{\ell}$, the $Z'_{X}$ model can make the maximum value of $Br(\tau\rightarrow\mu\nu_{\ell}\nu_{\ell})$ reach $1.1\times10^{-7}$. All $Z'$ models considered in this paper can produce significant contributions to the process $\tau\rightarrow\mu P$. However, the values of the branching ratio $Br(\tau\rightarrow\mu P)$ are still lower than the corresponding  experimental upper bounds. The value of the branching ratio $Br(P\rightarrow\mu e)$ is far below its corresponding experimental upper bound for all of the $Z'$ models.

 The extra gauge boson $Z'$, which can induce the  LFV couplings to ordinary leptons, might produce observable LFV signatures at the LHC. In the context of G(221) model, Ref.[18] has studied the possible signatures of the LFV couplings $Z'\ell_{i}\ell_{j}$  at the LHC and shown that, under reasonable expectations and conditions, the $e \mu $ signal could be used to test this NP model in near future. If one considers  the constraints of the experimental upper bound for the LFV  process  $\mu\rightarrow3e$ on the LFV coupling $Z'\mu e$, the production cross section and the number of $\mu e$ events will be significantly reduced. The final state with a lepton $\tau$ is  difficult to reconstruct from its decay products. However, the number of $\tau \mu$ or  $\tau e$ events is larger than the number of $\mu e$ events at least by three  orders of magnitude. This case is helpful to test the $Z'$ models, which will be carefully studied in near future.

\section*{Acknowledgments} \hspace{5mm}This work was
supported in part by the National Natural Science Foundation of
China under Grant No. 11275088, the Natural Science Foundation of the Liaoning Scientific Committee
(No. 2014020151).

\end{document}